\documentclass[11pt,a4paper]{article}
\usepackage{jcappub}
\usepackage{float}
\usepackage{color}

\newcommand{\eq}{\!\! =\!\!}

\DeclareSymbolFont{extraup}{U}{zavm}{m}{n}
\DeclareMathSymbol{\varheart}{\mathalpha}{extraup}{86}

\makeatletter
\newcommand{\vast}{\bBigg@{4}}
\newcommand{\Vast}{\bBigg@{6}}
\makeatother

\newcommand{\beq}{\begin{equation}}
\newcommand{\eeq}{\end{equation}}
\newcommand{\beqa}{\begin{eqnarray}}
\newcommand{\eeqa}{\end{eqnarray}}

\title{\Huge Cosmological Perturbation in $f(T)$ Gravity Revisited}

\author[\spadesuit]{\Large Keisuke Izumi,}
\author[\varheart, \spadesuit]{\Large Yen Chin Ong}
\affiliation[\spadesuit]{\large Leung Center for Cosmology and Particle Astrophysics,\\
 National Taiwan University, Taipei 10617, Taiwan}
\affiliation[\varheart]{\large Graduate Institute of Astrophysics,\\ National Taiwan University, Taipei 10617, Taiwan}
  
\emailAdd{izumi@phys.ntu.edu.tw}
\emailAdd{ongyenchin@member.ams.org}

\abstract{We perform detailed investigation of cosmological perturbations in $f(T)$ theory of gravity coupled with scalar field. Our work emphasizes on 
the way to gauge fix the theory and we examine all possible modes of perturbations up to \emph{second order}.
The analysis includes pseudoscalar and pseudovector modes in addition to the usual scalar, vector, and tensor modes. We find no 
gravitational propagating degree of freedom in the scalar, pseudoscalar, vector, as well as pseudovector modes. In addition, we find that the scalar and tensor perturbations have exactly the same form as their counterparts in usual general relativity with scalar field, except that the factor of reduced Planck mass squared $M_{\text{pl}}^2 \equiv 1/(8\pi G)$ that occurs in the latter has now been replaced by an effective \emph{time-dependent} gravitational coupling $-2 (df/dT)|_{T=T_0}$, with $T_0$ being the background torsion scalar. The absence of extra degrees of freedom of $f(T)$ gravity at second order linear perturbation indicates that $f(T)$ gravity is highly nonlinear. Consequently one cannot conclusively analyze stability of the theory without performing nonlinear analysis that can reveal the propagation of the extra degrees of freedom.}

\date{\today}

\begin{document}

\maketitle

\section{Introduction}

Einstein proposed the idea of teleparallelism, or \textit{Fernparallelismus}, to unify gravity and electromagnetism into a unified field theory in 1928 \cite{Einstein}. Unlike general relativity (GR) in which the Levi-Civita connection gives rise to curvature but vanishing torsion, in teleparallelism spacetime is endowed with a connection with vanishing curvature, but nonzero torsion. Since the curvature is identically zero, parallel transport of a vector is \textit{independent of path}. This is the origin of the name \textit{tele}parallel, which means ``parallel at a distance'' (Einstein's quest for unified field theory via teleparallelism is an interesting history and can be found in \cite{Sauer}). It has since been established that general relativity can in fact be re-cast into teleparallel language \cite{Hayashi, Pereira, Kleinert, Sonester}, known as the \textit{Teleparallel Equivalent of General Relativity} (TEGR). For an interesting formulation of TEGR as \emph{higher gauge theory}, see \cite{Baez}. 
Due to the need to understand the acceleration of the universe, various theories of modified gravity have been introduced, among which is the attempt to generalize TEGR to $f(T)$ theory of modified gravity in the same spirit as generalizing general relativity to $f(R)$ gravity \cite{0, 1}.

Given the spatially flat FLRW (Friedmann-Lema\^itre-Robertson-Walker) solution with suitable choice of the scale factor that describes an accelerating universe, 
the next step is to understand the evolution of structures on the FLRW background. 
This has been discussed from the observational point of view in \cite{Li:2011wu}.
On the theoretical level, perturbation theory is a useful tool for understanding the evolution of structures as it can reveal some properties of the dynamical modes of a gravitational theory. In the case of $f(T)$ theory, Li et. al \cite{Li} have shown  that there are \emph{generically}, 5 degrees of freedom in $f(T)$ gravity. Comparing with GR, which has only 2 degrees of freedom, there are 3 extra degrees of freedom, which the authors suggest could correspond to either one massive vector field or one
massless vector field together with one scalar field. 
Due to the high symmetry of FLRW metric,  there is \emph{no} extra degree of freedom \emph{at the background level}. This corresponds to the fact that the equations to solve for the background have the \emph{same} number of initial conditions as those of GR. Note that from the Hamiltonian perspective FLRW geometry has no dynamical degrees of freedom since the two degrees of freedom of gravitational waves are not excited in an isotropic universe. Indeed the only dynamics for FLRW universe is its expansion (or contraction). That is, the dynamics is completely determined by the Hubble parameter. This is the case in both GR and $f(T)$ gravity.

The absence of extra degrees on freedom in FLRW background is actually to be expected: When restricted to some particular geometry such as FLRW geometry or spherical geometry, some degrees of freedom must vanish \emph{by symmetry}. In particular spatial isotropy would not allow a non-vanishing spatial vector field, which selects a special direction. At the level of perturbation however, one must carefully check if any more degrees of freedom manifests themselves. In the case of $f(T)$ gravity, as pointed out in \cite{Li}, a number of extra degrees of freedom do not appear in Minkowski spacetime, even at linear perturbation level.

We remark that extra propagating degrees of freedom may imply additional interaction between two bodies. 
In order for the theory to be consistent with observational data, we need some mechanisms to suppress such interaction or nonlinear effects that prevent these extra modes from propagating. If $f(T)$ gravity is the exact theory describing our universe, then the FLRW solution should be stable. However, stability can only be established after we know the behaviors of all the degrees of freedom of $f(T)$. It is therefore an interesting question to ask: If we consider only linear perturbation, can we see \emph{any} extra degree of freedom propagating in the background of highly symmetric FLRW spacetime with flat spatial section? 

In order to address this question, we analyze the perturbative properties of $f(T)$ gravity on FLRW background.
Although linear perturbations on this particular background have been previously investigated in various papers \cite{Dent, Dent:2011zz, Cai:2011tc, Wu:2011kh, Wu:2012hs},
there remains room for improvement since the authors either ignored some equations or performed decomposition of degrees of freedom incompletely.
We believe that it is thus worthwhile to revisit the problem considering the most general type of perturbation, including all possible modes: scalar, pseudoscalar, vector, pseudovector, and tensor, and  by emphasizing on the way to gauge fix the theory. Furthermore, we perform our analysis up to \emph{second order} in perturbation, which is important for checking the ghost condition. 

The rest of this paper is organized as follows. For the sake of completeness, in Sec.~\ref{review} 
we shall review Teleparallel Equivalent of General Relativity and $f(T)$ theory. Our analysis starts in Sec.~\ref{analysis}. We show the exact way to decompose degrees of freedom  
in Subsec.~\ref{dec}, followed by the analysis of 
the second order action in each mode from Subsec.~\ref{scalar} to Subsec.~\ref{vector}. 
We conclude with some discussions in Sec.~\ref{Sum}. 

\section{From Teleparallel Equivalent of General Relativity to $f(T)$: A Review}
\label{review}

As in general relativity, the underlying structure of TEGR is a spacetime manifold $M$. At every point $p \in M$ in the local coordinate chart $\left\{x^1,\cdots, x^n\right\}$, the tangent space $T_pM$ at $p$ is spanned by the coordinate vector fields $\left\{\partial_1,\cdots,\partial_n\right\}$. The set $\left\{dx^1,\cdots,dx^n\right\}$ forms the basis for the dual space $T_p^*M$. Specializing to 4 dimensions, the tangent space is equipped with Lorentzian metric $g$ with signature $(-,+,+,+)$. We will label all spacetime coordinates by Greek indices that run from 0 to 3, with 0 denoting the time dimension, 
while all spatial coordinates will be labeled by $i,j,k,\ldots$ that run from 1 to 3. 

Let $\left\{e_A(x)\right\}$ be an arbitrary basis of $T_pM$. We can express the total covariant derivative $\nabla e_A$ as:
\begin{equation}
\nabla e_A (x) = \Gamma_{~A}^{B} (x) e_B(x),
\end{equation}
where the coefficient is the connection 1-form, satisfying
\begin{equation}
\Gamma_{~A}^B (x) = \left\langle \vartheta^B, \nabla e_A \right\rangle = \Gamma^B_{~\mu A} (x) dx^\mu,
\end{equation}
where $\left\{\vartheta^B(x)\right\}$ is the dual frame or \emph{coframe} of $\left\{e_B(x)\right\}$ and $\left\{dx^\mu\right\}$ is the dual basis of coordinate basis $\left\{\partial_\mu\right\}$. That is, $e_A = e^\mu_{~A} \partial_\mu$ and $\vartheta^A = e^A_{~\mu} dx^\mu$. In this work, $A,B,C,...$ take the values $0,1,2,3$ while $a,b,c,...$ take the value $1,2,3$.

Assuming that spacetime is parallelizable (i.e. there exist $n$ vector fields $\left\{v_1,...,v_n\right\}$ such that at \emph{any} point $p \in M$ the tangent vectors $v_i|_p$'s provide a basis of the tangent space at $p$), the mapping between the bases in coordinate frame $\left\{\partial_\mu\right\}$ to that of non-coordinate frame $\left\{e_A\right\}$ is an isomorphism $TM \to M \times \Bbb{R}^4$. This follows from the fact that any parallelizable $n$-dimensional manifold has a trivializable tangent bundle (in other words, $TM$ can be written as the direct product of $M$ and $\Bbb{R}^n$). Note that the frame field depends only on the affine structure of the manifold and hence \textit{a priori} has nothing to do with the metric. We now relate the frames with the metric by equipping the space $\Bbb{R}^4$ with a flat Minkowski metric $\eta_{_{AB}}$ such that 
\begin{equation}
g_{\mu\nu}=\eta_{_{AB}} e^A_{~\mu} e^B_{~\nu}.
\end{equation}
In other words, on the manifold $M$, we choose \emph{arbitrarily} a frame $e_A$ at each point, locally on some open chart $U \subset M$. By parallelizability, this can be extended globally. We then define a metric $\eta$ on $U$ by
\begin{equation}
\eta(e_{A},e_{B})=\eta_{_{AB}}.
\end{equation}
This imposes orthonormality on the tetrads. 

We remark that for gravity in (3+1)-dimensions, things work out nicely because Steenrod's theorem guarantees that all orientable 3-manifolds are parallelizable \cite{Steenrod, Stiefel}. Thus a 4-dimensional spacetime with orientable spatial section is also parallelizable. To put it slightly differently, if any spatial slice of spacetime is an orientable 3-manifold (and as such parallelizable) with initial data that can be propagated uniquely in time, in the manner of 3+1 decomposition of ADM \cite{ADM}, then the \emph{entire spacetime} is parallelizable. We remark also that a non-compact 4-dimensional Lorentzian manifold $M$ admits a spin structure if and only if it is parallelizable, a result known as Geroch's theorem \cite{Geroch}. 

One can now introduce the \emph{Weitzenb\"ock connection} \cite{Weitzenbock, Weitzenbock2} $\overset{w}\nabla$ defined by 
\begin{equation}
\overset{w}\nabla_X Y  := (XY^A)e_A,
\end{equation}
where $Y=Y^A e_A$. That is, we require that the tetrads are \emph{teleparallel}, i.e. covariantly constant: $\overset{w}\nabla_X e_A = 0$. This amounts to defining the connection coefficient by 
\begin{equation}
\overset{w}\Gamma~^\lambda_{~\mu \nu} = e_A^{~\lambda}\partial_\nu e^A_{~\mu} = - e^A_{~\mu} \partial_\nu e_A^{~\lambda}.
\end{equation}
Given any vector field $X$, there always exists such connection that gives rise to parallelization on $M$, assuming of course $M$ is parallelizable. Furthermore, the connection is unique \cite{Youssef}. 

The torsion tensor is defined by
\begin{flalign}
\overset{w}T(X,Y)& = \overset{w}\nabla_X Y - \overset{w}\nabla_Y X - [X,Y]\\
& = X^AY^B[e_A,e_B].
\end{flalign}
This expression is generically not zero since the basis vectors $\left\{e_A\right\}$ are not integrable in general. In local coordinates, the components of the torsion tensor are
\begin{equation}
\overset{w}T{}^\lambda{}_{\mu\nu} := \overset{w}\Gamma{}^\lambda_{~\nu\mu} - \overset{w}\Gamma{}^\lambda{}_{\mu\nu}=e_A^{~\lambda} (\partial_\mu e^A_{~\nu} - \partial_\nu{}e^A_{~\mu})\neq 0.
\end{equation}
 
The Weitzenb\"ock connection is curvature-free. One can verify this either by direct (but tedious) calculation in coordinates, or simply by noting that the definition of Riemann curvature endomorphism
\begin{equation}
\overset{w}R(X,Y)Z = \left(\overset{w}\nabla_X\overset{w}\nabla_Y-\overset{w}\nabla_Y\overset{w}\nabla_X-\overset{w}\nabla_{[X,Y]}\right)Z,
\end{equation}
and $\overset{w}\nabla_X e_A = (X \delta^C_A)e_C = 0$ together implies
\begin{equation}\label{Rc}
\overset{w}R(e_A,e_B)e_C = \overset{w}\nabla_{e_A} (\overset{w}\nabla_{e_B} e_C) - \overset{w}\nabla_{e_B} (\overset{w}\nabla_{e_A} e_C) - \overset{w}\nabla_{[e_A,e_B]}e_C.
\end{equation}
Every term of Eq.(\ref{Rc}) vanishes identically and so the Riemann curvature tensor is equal to zero. We would like to emphasize that curvature is not a property of the manifold itself, but depends on the choice of connection.  

The difference between the Weitzenb\"ock connection and the Levi-Civita connection is the \textit{contortion
 tensor}\footnote{Note the proper term is \emph{contortion} instead of contorsion. See e.g. \cite{Hehl}.} 
\begin{equation}\label{contortion0}
\overset{w}K{}^\lambda{}_{\mu\nu} :=  \frac{1}{2}\left(\overset{w}T_\nu{}^\lambda{}_\mu  + \overset{w}T_\mu{}^\lambda{}_\nu - \overset{w}T{}^\lambda{}_{\mu\nu}\right),
\end{equation}
or equivalently,
\begin{equation}\label{contortion}
\overset{w}K{}^{\mu\nu}{}_\rho=-\frac{1}{2}\left(\overset{w}T{}^{\mu\nu}{}_\rho - \overset{w}T{}^{\nu\mu}{}_\rho - \overset{w}T_\rho{}^{\mu\nu}\right).
\end{equation}

The Einstein-Hilbert action of general relativity (here the Ricci scalar $R$ is with respect to the Levi-Civita connection)
\begin{equation}\label{gr}
S = \frac{1}{2\kappa} \int{ d^4x \sqrt{-g} R} , ~~~ \kappa = 8\pi G,
\end{equation}
can be written as
\begin{equation}\label{TEGR}
S=-\frac{1}{2\kappa} \int{d^4x~e \overset{w}T},
\end{equation}
where
$e=|\det(e^A_{~\mu})|$, which is equal to $\sqrt{-g}$ in GR, and

\begin{equation}\label{T}
\overset{w}T := \overset{w}S_\rho{}^{\mu\nu}\overset{w}T{}^\rho{}_{\mu\nu},
\end{equation}
with
\begin{equation}
\overset{w}S_\rho{}^{\mu\nu}\equiv \frac{1}{2} \left(\overset{w}K{}^{\mu\nu}{}_\rho + \delta^\mu_\rho \overset{w}T{}^{\alpha\nu}{}_{\alpha} - \delta^\nu_\rho \overset{w}T{}^{\alpha \mu}{}_{\alpha} \right).
\end{equation}
The quantity $\overset{w}{T}$ is the so-called \emph{torsion scalar}. 
Explicitly, it is written as 
\begin{equation}\label{torsion scalar}
\overset{w}{T} = \frac{1}{4}\overset{w}T{}^\rho{}_{\eta\mu}\overset{w}T_\rho{}^{\eta\mu} + \frac{1}{2}\overset{w}T{}^\rho{}_{\mu\eta}\overset{w}T{}^{\eta\mu}{}_\rho - \overset{w}T_{\rho\mu}{}^\rho \overset{w}T{}^{\nu\mu}{}_\nu.
\end{equation}
The equivalence between Eq.(\ref{gr}) and Eq.(\ref{TEGR}) comes from the equality
\begin{eqnarray}
R=-\overset{w}T-2 \nabla^\mu \overset{w}{T}{}^\nu_{\ \mu\nu},\label{relation}
\end{eqnarray}
where $\nabla$ is the covariant derivative in GR.

The Einstein-Hilbert action in the form of Eq.(\ref{TEGR}) is known as the \textit{TEGR action}. $f(T)$ gravity is obtained by replacing $T$ in TEGR action with a function $f(T)$. 

Note that unlike TEGR, we \emph{do not} have local Lorentz invariance in $f(T)$ gravity, 
because the total derivative term $\nabla^\mu \overset{w}{T}{}^\nu_{\ \mu\nu}$ in Eq.(\ref{relation}) breaks local Lorentz invariance
(see e.g. discussions in \cite{Barrow1} and \cite{Barrow2}). Unlike in general relativity in which we can change coordinate systems and use any frame fields, we have no such freedom in $f(T)$ gravity. In particular, this implies that we cannot directly extract the tetrad from the metric in a straightforward manner. TEGR is special since its action does not determine the admissible frame but only the metric. Nevertheless, in the FLRW cosmology with \emph{flat} spatial section, 
\begin{eqnarray}
&&ds^2 = g_{\mu\nu}dx^\mu dx^\nu= -dt^2 + a^2 \delta_{ij}dx^i dx^j, \label{FLRW}
\end{eqnarray}
where $a=a(t)$ is the expansion factor,
the standard choice of tetrads, or equivalently the cotetrads, related to the coordinate one-forms by 
\begin{eqnarray}
&& e^0_{\ \mu}dx^\mu= dt,\\
&& e^a_{\ \mu}dx^\mu= a \delta_{ai} dx^i.
\end{eqnarray}
is a suitable choice of frame. However this tetrad is not the most general one. For example, one may wish to consider spacetime with less symmetry such as Bianchi models. In fact this will be quite important for the analysis of $f(T)$ gravity. We will further comment on this issue in Sec.~\ref{Sum}.

From now onwards, we will suppress all explicit overscript $w$'s on the torsion scalar and connection coefficients etc. 

\section{$f(T)$ Gravity and Its Cosmological Perturbation}
\label{analysis}

In this work, we consider the cotetrads
\begin{eqnarray}
&&\bar e^0_{\ \mu}dx^\mu= dt,\\
&&\bar e^a_{\ \mu}dx^\mu= a \delta_{ai} dx^i,
\end{eqnarray}
that is, the background metric is taken to be the usual FLRW geometry with flat spatial section given by Eq.(\ref{FLRW}). We will denote the Hubble parameter by $H \equiv \dot{a}/a$. Here the bar notation emphasizes the fact that the cotetrads are that of the background.

The gravitational action is\footnote{In our convention, the factor of $(2\kappa)^{-1}$ is absorbed into $f$.} 
\begin{eqnarray}\label{f(t)}
S_g=\int d^4 x \ e f(T).
\end{eqnarray}

We also introduce a scalar field into our theory as a toy model for the matter sector. The matter scalar field is described by the action
\begin{eqnarray}
S_m= \int d^4 x \  e 
\left( -\frac{1}{2}g^{\mu\nu}\partial_\mu \phi\partial_\nu \phi -V(\phi) \right),
\end{eqnarray}
where $V(\phi)$ is an unspecified potential.

\subsection{Decomposition of Degrees of Freedom and Gauge Freedom}
\label{dec}

Using the background symmetry, we can decompose the perturbative cotetrad into five scalars, one pseudoscalar, three vectors, one pseudovector and one tensor%
\footnote{
In \cite{Wu:2012hs}, the authors had discussed the decomposition but it is imperfect. 
Although they had tried to write the antisymmetric part of $\delta e^a_{\ i}$ in terms of vector and scalar, 
this should more appropriately be written in terms of \emph{pseudoscalar} and \emph{pseudovector}. 
While pseudovector can be written in terms of vector due to the relation Eq.(\ref{coupling}),
pseudoscalar can never be written in terms of scalar. 
}. In the following calculations, all equalities are understood to be either exact, or valid up to second order only.

The exact decomposition is
\begin{eqnarray}
&&\delta e^0_{\ t} = \Phi,\\
&&\delta e^0_{\ i} = a \left( \partial_i \beta + u_i\right),\\ 
&&\delta e^a_{\ t} = \delta_{ai} \left( \partial_i B + v_i \right),\\
&&\delta e^a_{\ i} = a \delta_{aj} \left[ \delta_{ij} \psi +\partial_i \partial_j E
+\partial_i w_j +\partial_j w_i +h_{ij} + 
\epsilon_{ijk} \left( \partial_k \tilde \sigma +\tilde V_k\right)\right],
\end{eqnarray}
where $\Phi$, $\beta$, $B$, $\psi$ and $E$ are scalars, $\tilde \sigma$ is a pseudoscalar, $u_i$, $v_i$ and $w_i$ are vectors, $\tilde V_i$ is a pseudovector and $h_{ij}$ is a
tensor. We note that vectors and pseudovectors can be coupled with each other in the following form:
\begin{eqnarray}
\epsilon_{ijk} \left( \partial_i u_j \right) \tilde V_k. \label{coupling}
\end{eqnarray}
If the theory preserves parity, no other modes of decomposition can be coupled with each other. This means that we can, with the exception of vector and pseudovector modes, study the different modes separately. 

As in GR, the action of $f(T)$ gravity is invariant under coordinate transformation:
\begin{eqnarray}
x^\mu \to {x'}^\mu = x^\mu + \xi^\mu(x).
\end{eqnarray}
Under this transformation, the cotetrad transforms as
\begin{eqnarray}
{e'}^A_{\ \mu} (x') = e^A_{\ \nu} \frac{\partial x^\nu}{\partial {x'}^\mu}.
\end{eqnarray}
This leads to the gauge transformation of perturbative variables on FLRW metric (\ref{FLRW}) as 
\begin{eqnarray}
&&\Phi \to \Phi' =\Phi-\dot \xi^t.\\
&&\beta\to\beta'= \beta - \frac{1}{a} \xi^t,\\
&&B\to B' =B -\left(\dot \xi -\frac{\dot a}{a} \xi\right),\label{B}\\
&&\psi\to\psi'=\psi-\frac{\dot a}{a} \xi^t,\\
&&E\to E' =E-\frac{1}{a} \xi,\label{E}\\
&&\tilde \sigma \to \tilde \sigma' =\tilde \sigma, \\
&&u_i \to u'_i =u'_i,\\
&&v_i \to v'^i =v^i -\left(\dot \xi^{(v)}_i -\frac{\dot a}{a}\xi^{(v)}_i\right), \\
&&w_i \to w'_i =w_i -\frac{1}{2a} \xi^{(v)}_i,\\
&&\tilde V_i \to \tilde V'_i =\tilde V_i- \frac{1}{a} \epsilon_{ijk} \partial_j \xi^{(v)}_k,\\
&&h_{ij} \to h'_{ij}=h_{ij},
\end{eqnarray}
where we have decomposed $\xi^i$ as $\xi^i = a^{-1} (\partial_i \xi + \xi^{(v)}_i)$ 
and dot denotes the partial derivative with respect to time. 
We can choose the gauge in which one element from each of the sets $\left\{\Phi, \beta, \psi\right\}$, $\left\{B, E\right\}$, 
and $\left\{v_i, w_i, \tilde V_i\right\}$ vanishes. That is, by gauge-fixing, the pseudovector mode $\tilde V_i$ can be set to zero and thus we only need to consider perturbative analysis of vector modes. We now begin our perturbative analysis of the various modes of degrees of freedom, starting from scalar perturbation.

\subsection{Scalar Perturbation}
\label{scalar}

To carry out scalar perturbation, we shall first choose the gauge where $\beta=B=0$. 
For the sake of convenience, we shall introduce another variable $\zeta$, which is defined as 
\begin{eqnarray}
\zeta \equiv \psi+ \frac{\Delta E}{3}, \label{defzeta}
\end{eqnarray}
where $\Delta \equiv \delta_{ij} \partial _i \partial_j$. We will use this new variable in place of $\psi$.

The cotetrad and metric can now be written as 
\begin{eqnarray}
&&e^0_{\ \mu}dx^\mu= \left(1+\Phi\right) dt,\\ 
&&e^a_{\ \mu} dx^\mu =\delta_{ai}
\left\{a \left[\delta_{ij}(1+ \zeta)+ \left(\partial_i \partial_j-\frac{1}{3}
\Delta\delta_{ij} \right) E\right] \right\} dx^j,\\
&&e_0^{\ \mu} \frac{\partial}{\partial x^\mu}=  
(1-\Phi+\Phi^2) \frac{\partial}{\partial t},\\
&&e_a^{\ \mu}\frac{\partial}{\partial x^\mu}= 
a^{-1} \biggl\{\delta _{ai} (1- \zeta +\zeta^2 ) - \delta _{aj}\left(\partial_i \partial_j -\frac{1}{3}\Delta \delta_{ij}\right)E 
+2 \delta_{aj} \zeta \left(\partial_i \partial_j -\frac{1}{3}\Delta \delta_{ij}\right)E\nonumber\\
&&\qquad\qquad\qquad\qquad
+\delta_{aj} \left[\left(\partial_i \partial_k -\frac{1}{3}\Delta \delta_{ik}\right)E\right]
\left[\left(\partial_j \partial_k -\frac{1}{3}\Delta \delta_{jk}\right)E\right] \biggr\}
\frac{\partial}{\partial x^i},\\
&&g_{\mu\nu}dx^\mu dx^\nu= - (1+\Phi)^2dt^2 \nonumber \\
&&\qquad\qquad\qquad+ 
a^2 \biggl\{\delta_{ij} (1+\zeta)^2 +2 (1+\zeta)\left(\partial_i  \partial_j  -\frac{1}{3}\Delta\delta_{ij} \right) E \nonumber \\
&&\qquad\qquad\qquad\qquad
+\left[\left(\partial_i  \partial_k  -\frac{1}{3}\Delta\delta_{ik} \right) E\right]
\left[\left(\partial_j  \partial_k  -\frac{1}{3}\Delta\delta_{jk} \right) E\right]\biggr\}
dx^i dx^j,\\
&&g^{\mu\nu}\frac{\partial}{\partial x^\mu}\frac{\partial}{\partial x^\nu}=
-(1-2\Phi+3\Phi^2) \frac{\partial}{\partial t}\frac{\partial}{\partial t}\nonumber \\
&&\qquad\qquad\qquad\qquad
+a^{-2}
\biggl\{\delta_{ij}-2\delta_{ij}\zeta
-2 \left(\partial_i \partial_j-\frac{1}{3}\Delta\delta_{ij}\right)E
+3\delta_{ij}\zeta^2 +6\zeta \left(\partial_i\partial_j -\frac{1}{3}\Delta\delta_{ij}\right)E\nonumber \\
&&\qquad\qquad\qquad\qquad
+3\left[\left(\partial_i\partial_k -\frac{1}{3}\Delta\delta_{ik}\right)E\right]
\left[\left(\partial_j\partial_k -\frac{1}{3}\Delta\delta_{jk}\right)E\right]
\biggr\}
\frac{\partial}{\partial x^i}\frac{\partial}{\partial x^j}.
\end{eqnarray}

Then, we have
\begin{equation}
e = a^3 \left[
1+\Phi + 3 \zeta +3 \Phi \zeta +3\zeta^2 +\frac{1}{6} (\Delta E)^2 -\frac{1}{2}
(\partial_i \partial_j E)(\partial_i \partial_j E)
\right].
\end{equation}

The components of the torsion tensors are
\begin{eqnarray}
&&T^t_{\ ti}= -\partial_i \Phi + \Phi \partial_i \Phi, \\
&&T^t_{\ ij}= 0, \\
&&T^i_{\ tj} = H \delta_{ij}+ \delta_{ij} \dot \zeta + 
\left(\partial_i \partial_j -\frac{1}{3}\Delta \delta_{ij}\right)\dot E 
-(\partial_i \partial_k E)\left[
\left(\partial_k \partial_j -\frac{1}{3}\Delta \delta_{kj}\right)\dot E\right]\nonumber\\
&&\qquad\qquad
-\dot \zeta (\partial_i \partial_j E)
+\left(- \zeta +\frac{1}{3}\Delta E\right)
\left[\delta_{ij} \dot \zeta 
+\left(\partial_i \partial_j -\frac{1}{3}\Delta \delta_{ij}\right)\dot E\right]
, \\
&&T^i_{\ jk} = \partial_j \delta_{ik} \zeta -\frac{1}{3}\Delta \partial_j \delta_{ik} E
+\left(-\zeta +\frac{1}{3} \Delta E\right)
\left(\partial_j \delta_{ik} \zeta -\frac{1}{3}\Delta \partial_j \delta_{ik} E\right)
+ \frac{1}{3}(\partial_i \partial_k E) (\Delta \partial_j E) \nonumber\\
&&\qquad\qquad
-(\partial_i \partial_k E)(\partial_j \zeta)- (j \leftrightarrow k), \label{ijk}\\
&&T^\mu_{\ t \mu} = 3H + 3\dot \zeta -3 \zeta \dot \zeta 
-(\partial_i \partial_j E)(\partial_i \partial_j \dot E) +\frac{1}{3}(\Delta E)(\Delta \dot E),\\
&&T^\mu_{\ i\mu}= \partial_i \Phi +2 \partial_i \zeta -\frac{2}{3}\Delta\partial_i E 
- \Phi \partial_i \Phi +\left( -\zeta +\frac{1}{3} \Delta E\right)\left(2 \partial_i \zeta -\frac{2}{3}\Delta\partial_i E\right)\nonumber\\
&&\qquad\qquad
+\frac{1}{3}(\Delta E)(\Delta \partial_i E) - \frac{1}{3} (\partial_i \partial_j E)(\partial_j \Delta E)-(\Delta E)(\partial_i \zeta)
+(\partial_j \partial_i E)(\partial_j \zeta),
\end{eqnarray}
where $(j \leftrightarrow k)$ in Eq.(\ref{ijk}) denotes the permutations of the all previous terms with respect to $j$ and $k$.

Define $T_\mu\equiv T^\nu_{\ \mu\nu}$. Then, we have the following contractions:

\begin{eqnarray}
&&T_\mu T_\nu g^{\mu\nu}= -9H^2 -18H \dot \zeta +18 H^2 \Phi -9 \dot\zeta ^2 +
36 H \Phi \dot \zeta -27 H^2 \Phi^2 +18H \zeta \dot \zeta \nonumber\\
&&\qquad
+6H(\partial_i\partial_j E)(\partial_i\partial_j \dot E) -2(\Delta E)(\Delta \dot E)H
+ a^{-2} \left[\partial_i \left(\Phi+ 2\zeta -\frac{2}{3}\Delta E\right)\right]^2, \label{1}\\
&&\frac{1}{4}T^\rho_{\ \mu\nu}T^\alpha_{\ \beta\gamma}g_{\rho\alpha}g^{\mu\beta}g^{\nu\gamma}
+\frac{1}{2}T^\rho_{\ \mu\nu}T^\nu_{\ \alpha\rho}g^{\mu\alpha} \nonumber \\
&&\qquad
= -3H^2 -6H \dot \zeta +6H^2 \Phi + a^{-2} (\partial_i  \Phi)^2
+2 a^{-2} \left[\partial_i \left(\zeta-\frac{1}{3}\Delta E\right) \right]^2\nonumber \\
&&\qquad\qquad
-9H^2\Phi^2 +12H\Phi \dot \zeta-3 \dot \zeta^2 +6H\zeta \dot \zeta 
-\frac{2}{3}H(\Delta E)(\Delta \dot E) \nonumber \\
&&\qquad\qquad
+2H(\partial_i \partial_j E)(\partial_i \partial_j \dot E)
-(\partial_i \partial_j \dot E)^2+\frac{1}{3}(\Delta \dot E)^2. \label{2}
\end{eqnarray}

The difference between Eq.(\ref{2}) and Eq.(\ref{1}) gives precisely the torsion scalar (see Eq.(\ref{torsion scalar})) up to second order in perturbation:
\begin{eqnarray}
&&T= 6H^2 +12 H \dot \zeta -12 H^2 \Phi 
-4 a^{-2} (\partial_i \Phi )\left[\partial_i\left(\zeta -\frac{1}{3}\Delta E\right)\right]
-2a^{-2}\left[\partial_i\left(\zeta -\frac{1}{3}\Delta E\right)\right]^2\nonumber \\
&&\qquad
+18 H^2 \Phi^2 -24 H \Phi \dot \zeta + 6 \dot \zeta^2 -12 H \zeta \dot \zeta
-4 H (\partial_i \partial_j E)(\partial_i \partial_j \dot E)\nonumber \\
&&\qquad
+\frac{4}{3}H(\Delta E)(\Delta \dot E)
-(\partial_i \partial_j \dot E)^2 + \frac{1}{3}(\Delta \dot E)^2.
\end{eqnarray}

Note that $T_0 \equiv 6H^2$ is simply the background torsion scalar for FLRW geometry with flat spatial section. Most calculations in literature have the torsion scalar to be $-6H^2$ instead. This difference by a minus sign can be directly checked to be the result of our choice of east coast sign convention $(-,+,+,+)$ instead of west coast one $(+,-,-,-)$.  

We can now expand the gravitational part of the action. Let us denote $f_0 = f(T_0)$ and let prime denotes the derivative $(df/dT)|_{T=T_0}$. Then, upon integrating by parts and dropping the surface term, we have 

\begin{eqnarray}
S_g^S
&\eq&\int dt d^3x ~a^3 \biggl[ f_0 + (f_0-12H^2 f_0') \Phi 
+3(f_0 -12H^2 f_0'- 4 \dot Hf_0'-48H^2 \dot Hf_0'')\zeta  \nonumber\\
&&\qquad\qquad\qquad
+6H^2 (f_0' +12H^2 f_0'') \Phi^2 -12 H (f_0'+12H^2 f_0'')\Phi \dot \zeta
+3(f_0-12H^2f_0')\Phi \zeta \nonumber\\
&&\qquad\qquad\qquad
+ 4 a^{-2} f_0' \Phi \Delta \left(\zeta -\frac{1}{3} \Delta E\right)
+6(f_0'+12H^2f_0'') \dot \zeta^2 + 24 H f_0' \zeta \dot \zeta + 3f_0 \zeta^2 \nonumber\\
&&\qquad\qquad\qquad
-\frac{2}{3} f_0'(\Delta \dot E)^2 -\frac{8}{3}Hf_0'(\Delta E)(\Delta \dot E)
-\frac{1}{3}f_0(\Delta E)^2\nonumber\\
&&\qquad\qquad\qquad 
+ 2a^{-2} f_0' \left(\zeta -\frac{1}{3}\Delta E\right) \Delta \left(\zeta -\frac{1}{3}\Delta E\right) \biggr].
\end{eqnarray}

We must also expand the action for scalar field. 
We define the perturbation of scalar field as 
\begin{eqnarray}
\phi = \phi_0 +\delta \phi,
\end{eqnarray}
where $\phi_0$ is the background value of $\phi$ and $\delta\phi$ is the perturbation of $\phi$.
Then, the scalar field action can be written up to the second order of perturbation as 
\begin{eqnarray}
&&S_m^S= \int d^4 x \ a^3 \biggl[\frac{1}{2}\dot \phi_0^2 -V 
+ \left(\frac{1}{2}\dot \phi_0^2-V\right)(\Phi +3\zeta) + \dot \phi_0 \dot{\delta \phi} 
-\dot \phi_0^2\Phi -V' \delta\phi\nonumber\\
&&\qquad+\left(\frac{1}{2} \dot \phi_0^2-V\right)\left(3 \Phi\zeta+3\zeta^2- \frac{2}{3}(\Delta E)^2\right)
+(\Phi +3\zeta) ( \dot \phi_0 \dot {\delta \phi} -\dot \phi_0^2 \Phi -V'\delta \phi )\nonumber\\
&&\qquad\qquad
-2\dot \phi_0 \dot {\delta \phi} \Phi + \frac{1}{2} \dot {\delta \phi}^2 
+ \frac{1}{2 a^2} \delta \phi \Delta\delta \phi - \frac{1}{2} V'' \delta\phi^2 
+ \frac{3}{2}\dot \phi_0^2 \Phi^2
\biggr].
\end{eqnarray}

From the variation of the first order action with respect to $\Phi$, $\zeta$ and $\delta\phi$, 
we can obtain the equations of motion of the background as

\begin{eqnarray}
&&f_0-12H^2 f_0'-\frac{1}{2}\dot \phi_0^2 -V=0, \label{Phi} \\
&&f_0 -12H^2 f_0'- 4 \dot Hf_0'-48H^2 \dot Hf_0'' +\frac{1}{2}\dot \phi_0^2 -V=0 , \label{zeta}\\
&&\ddot \phi_0 + 3H \dot \phi_0 +V'=0.  \label{phi}
\end{eqnarray}
The first two equations give 
\begin{equation}
\dot \phi_0^2 = 4\dot H(f_0'+ 12 H^2 f_0''). \label{dotphi}
\end{equation}

We now proceed to the second order action. It is now more convenient to stop using $\zeta$ and 
switch back to the original variable $\psi$. We recall that the relation between these two 
variables is given by Eq.(\ref{defzeta}).
Moreover, we shall change to another choice of gauge in which $E=\beta=0$, in place of the original gauge choice
$\beta=B=0$.  

The second order action in the new gauge is easily obtained by replacing $\dot E$ with $-a^{-1}B$ in the corresponding equations which were obtained using the original gauge choice
\footnote{
Note that $\dot E -a^{-1}B$ is an gauge invariant combination. See Eq.(\ref{B}) and Eq.(\ref{E}).
}
.
Then, integrating by parts, dropping the surface terms and 
using Eqs.(\ref{Phi})-(\ref{dotphi}),
the second order action can be written as  
\begin{eqnarray}
&&S_2^S=\int dt d^3x ~a^3\nonumber \\
&&\qquad \times
\Biggl[
\frac{\dot \phi_0^2}{2 \dot H}(\dot H +3 H^2)
\biggl\{\Phi + \frac{ \dot H} {\dot \phi_0^2(\dot H + 3H^2)} 
\biggl(- \frac{\dot \phi_0^2}{\dot H H}(\dot H +3H^2) \dot \psi
+\frac{\dot \phi_0^2}{H^2}(\dot H +3H^2)  \psi \nonumber\\
&&\qquad\qquad\qquad\qquad\qquad\qquad\qquad
+4a^{-2}f_0'\Delta\psi -\dot \phi_0 \dot \alpha- V'\alpha 
+\frac{H}{\dot H} \dot \phi_0^2 a^{-1}\Delta B\biggr)\biggr\}^2 \nonumber\\
&&\qquad\qquad
-\frac{8f_0'^2 \dot H}{\dot \phi_0^2 (\dot H + 3 H^2)}
\biggl\{
a^{-2} \Delta \psi -\frac{\dot \phi_0^2 (\dot H + 3 H^2)}{16f_0'^2 \dot H}
\biggl(\frac{4 \dot H f_0'}{\dot\phi_0(\dot H + 3H^2)}\dot \alpha \nonumber\\
&&\qquad\qquad\qquad\qquad\qquad\qquad
-\frac{4 H f_0'}{\dot H + 3H^2} a^{-1}\Delta B +\frac{\dot \phi_0}{H} \alpha
+\frac{4 \dot H V' f_0'}{\dot \phi_0^2 (\dot H + 3H^2)} \alpha
\biggr)
\biggr\}^2\nonumber\\
&&\qquad\qquad
+8H^2 f_0'' \left(a^{-1}\Delta B- \frac{3}{4f_0'}\dot\phi_0 \alpha \right)^2\nonumber\\
&&\qquad\qquad
+\frac{1}{2}\dot \alpha^2 +\frac{1}{2a^2} \alpha \Delta\alpha -\frac{1}{2}V''\alpha^2 
+\left(\frac{\dot \phi_0^4}{16f_0'^2H^2} + \frac{3 \dot \phi_0^2}{4 f_0'} 
+\frac{\dot \phi_0 V'}{2Hf_0'}\right) \alpha^2
\Biggr],
\label{scalarsecond}
\end{eqnarray}
where 
\begin{eqnarray}
\alpha \equiv \delta\phi - \frac{\dot \phi_0}{H} \psi.
\end{eqnarray}
We can see that there is no kinetic term for $\Phi$, $\psi$ or $\beta$ 
and so we have only a \emph{single} propagating degree of freedom%
\footnote{
In papers \cite{Dent, Dent:2011zz, Cai:2011tc, Wu:2011kh}, 
the authors fixed the three scalar modes by appealing to gauge degrees of freedom. This is an \emph{over-fixing}. 
There are only two scalar degrees of freedom in the gauge modes, and thus we can fix only two scalar modes by gauge fixing. 
An over-fixed gauge condition may lead to false result. 
For example, if we impose not only the gauge fixing condition $E=\beta=0$ 
but also another condition $\Phi=0$,  
we \emph{cannot} obtain the constraint equation for $\Phi$ and 
$(\partial_t \psi)$ squared that remains in the second order action in Eq.(\ref{scalarsecond}). 
This gives the false result that there are two propagating scalar degrees of freedom instead of just one.
}. 
Moreover, $\alpha$ is of the same form as the gauge invariant variable in GR. As a result, 
the only difference from the corresponding second order perturbative expansion in GR is
the replacement of $1/(2 \kappa)$ by an effective gravitational coupling $-f_0'$.

\subsection{Pseudoscalar Perturbation}

The cotetrads and metric for pseudoscalar perturbation are given by
 
\begin{eqnarray}
&&e^0_{\ \mu}dx^\mu= dt,\\ 
&&e^a_{\ \mu} dx^\mu =\delta_{ai}
a \left[\delta_{ij} + \epsilon _{ijk} \partial_k \tilde \sigma \right]  dx^j,\\
&&e_0^{\ \mu} \frac{\partial}{\partial x^\mu}= \frac{\partial}{\partial t},\\
&&e_a^{\ \mu}\frac{\partial}{\partial x^\mu}= 
\delta _{ai} a^{-1} \left( \delta_{ij} +\epsilon_{ijk} \partial_k \tilde\sigma 
-\delta_{ij} (\partial_k \tilde \sigma)^2 + (\partial_i \tilde \sigma)(\partial_j \tilde \sigma)
\right)
\frac{\partial}{\partial x^j},\\
&&g_{\mu\nu}dx^\mu dx^\nu= - dt^2 +a^2 \left(\delta_{ij} 
+\delta_{ij} (\partial_k \tilde \sigma)^2 
-(\partial_i \tilde \sigma)(\partial_j \tilde \sigma)\right)dx^i dx^j,\\
&&g^{\mu\nu}\frac{\partial}{\partial x^\mu}\frac{\partial}{\partial x^\nu}=
- \left(\frac{\partial}{\partial t}\right)^2+a^{-2}
\left(\delta_{ij} -\delta_{ij} (\partial_k \tilde \sigma)^2 
+(\partial_i \tilde \sigma)(\partial_j \tilde \sigma)
\right)\frac{\partial}{\partial x^i}\frac{\partial}{\partial x^j}.
\end{eqnarray}

This leads to
\begin{eqnarray}
&&e = a^3 \left(1+(\partial_i \tilde \sigma)^2\right),
\end{eqnarray}
and the following components of torsion tensors:
\begin{eqnarray}
&&T^t_{\ ti}= 0, \\
&&T^t_{\ ij}= 0, \\
&&T^i_{\ tj} = H \delta_{ij}+\epsilon_{ijk} \partial_k \dot {\tilde\sigma} 
+\delta_{ij} (\partial_k \tilde \sigma)(\partial_k \dot {\tilde \sigma})
-(\partial_j \tilde \sigma)(\partial_i \dot {\tilde \sigma}), \\
&&T^i_{\ jk} =\epsilon_{ikl} \partial_l \partial_j \tilde \sigma
-\epsilon_{ijl} \partial_l \partial_k \tilde \sigma 
+\delta_{ik} (\partial_l \tilde \sigma)(\partial_l \partial_j \tilde \sigma)
\nonumber\\
&&\qquad\qquad
-\delta_{ij} (\partial_l \tilde \sigma)(\partial_l \partial_k \tilde \sigma)
-(\partial_k \tilde \sigma)(\partial_i \partial_j \tilde \sigma)
+(\partial_j \tilde \sigma)(\partial_i \partial_k \tilde \sigma),\\
&&T^\mu_{\ t \mu} = 3H +2 (\partial_i \tilde \sigma)(\partial_i \dot {\tilde \sigma}),\\
&&T^\mu_{\ i\mu}= (\partial_j \tilde \sigma)(\partial_j \partial_i \tilde \sigma)
+(\partial_i \tilde \sigma)(\Delta \tilde \sigma).
\end{eqnarray}

We then compute the various contractions:
\begin{eqnarray}
&&T_\mu T_\nu g^{\mu\nu}= -9H^2 
-12H (\partial_i \tilde \sigma)(\partial_i \dot {\tilde \sigma}) , \\
&&\frac{1}{4}T^\rho_{\ \mu\nu}T^\alpha_{\ \beta\gamma}g_{\rho\alpha}g^{\mu\beta}g^{\nu\gamma}
+\frac{1}{2}T^\rho_{\ \mu\nu}T^\nu_{\ \alpha\rho}g^{\mu\alpha} \nonumber \\
&&\qquad
= -3H^2 -4H(\partial_i \tilde\sigma)(\partial_i \dot {\tilde \sigma}) 
+a^{-2} \left[(\partial_i\partial_j \tilde \sigma)^2 -(\Delta \tilde \sigma)^2 \right].
\end{eqnarray}

Finally, we can obtain the torsion scalar
\begin{eqnarray}
&&T= 6H^2 +8H(\partial_i \tilde\sigma)(\partial_i \dot {\tilde \sigma}) 
+a^{-2} \left[(\partial_i\partial_j \tilde \sigma)^2 -(\Delta \tilde \sigma)^2 \right].
\end{eqnarray}

The second order action for gravitational field is

\begin{eqnarray}
S_{g,2}^{PS}= \int dt d^3x \ a^3 
\left[ f (\partial_i \tilde \sigma)^2 
+8Hf'(\partial_i \tilde\sigma)(\partial_i \dot {\tilde \sigma}) 
+a^{-2} f' \left[(\partial_i\partial_j \tilde \sigma)^2 -(\Delta \tilde \sigma)^2 \right] \right].
\end{eqnarray}

On the other hand, the second order scalar field action is 
\begin{eqnarray}
S_{m,2}^{PS}= \int dt d^3x \ a^3 
\left[\left(\frac{1}{2} \dot \phi_0^2 -V\right)(\partial_i \tilde\sigma)^2\right].
\end{eqnarray}

As before, we integrate by parts and drop the surface terms. By the background equation Eq.(\ref{zeta}), 
the sum of the above two actions $S_{g,2}^{PS}$ and $S_{m,2}^{PS}$ is zero. Thus, we see that up to second order in perturbation, there is no propagating pseudoscalar degree of freedom.
 
\subsection{Tensor Perturbation}

Moving on to tensor perturbation, the cotetrads and metric are given by

\begin{eqnarray}
&&e^0_{\ \mu}dx^\mu= dt,\\ 
&&e^a_{\ \mu} dx^\mu =\delta_{ai}
\left(\delta_{ij}+h_{ij} \right) dx^j,\\
&&e_0^{\ \mu} \frac{\partial}{\partial x^\mu}=\frac{\partial}{\partial t},\\
&&e_a^{\ \mu}\frac{\partial}{\partial x^\mu}= 
a^{-1} \delta _{aj} (\delta_{ij}-h_{ij}+h_{ik}h_{kj})
\frac{\partial}{\partial x^i},\\
&&g_{\mu\nu}dx^\mu dx^\nu= - dt^2 +a^2 (\delta_{ij} +2h_{ij} +h_{ik}h_{kj})
dx^i dx^j,\\
&&g^{\mu\nu}\frac{\partial}{\partial x^\mu}\frac{\partial}{\partial x^\nu}=
-\left(\frac{\partial}{\partial t}\right)^2
+a^{-2}(\delta_{ij} -2h_{ij} +3h_{ik}h_{kj})\frac{\partial}{\partial x^i}\frac{\partial}{\partial x^j}.
\end{eqnarray}

We thus have
\begin{eqnarray}
&&e = a^3 \left(1-\frac{1}{2}h_{ij}^2\right).
\end{eqnarray}
The components of the torsion tensor are:
\begin{eqnarray}
&&T^t_{\ ti}= 0, \\
&&T^t_{\ ij}= 0, \\
&&T^i_{\ tj} = H \delta_{ij}+ \dot h_{ij} -h_{ik} \dot h_{kj}, \\
&&T^i_{\ jk} =\partial_j h_{ik}- \partial_k h_{ij}-h_{il}\partial_j h_{lk}
+h_{il} \partial_k h_{lj},\\
&&T^\mu_{\ t \mu} = 3H -h_{ij} \dot h_{ij},\\
&&T^\mu_{\ i\mu}= -h_{jk} \partial_i h_{jk}+ h_{jk} \partial_j h_{ki}.
\end{eqnarray}

We proceed to calculate in the same way as what we did before
\begin{eqnarray}
&&T_\mu T_\nu g^{\mu\nu}= -9H^2 +6H h_{ij} \dot h_{ij}, \\
&&\frac{1}{4}T^\rho_{\ \mu\nu}T^\alpha_{\ \beta\gamma}g_{\rho\alpha}g^{\mu\beta}g^{\nu\gamma}
+\frac{1}{2}T^\rho_{\ \mu\nu}T^\nu_{\ \alpha\rho}g^{\mu\alpha} \nonumber \\
&&\qquad
= -3H^2 - \dot h_{ij}^2 +2H h_{ij} \dot h_{ij} +a^{-2} \left[ (\partial_i h_{jk})^2 
-(\partial_i h_{jk})(\partial_j h_{ik})\right],
\end{eqnarray}
and obtain the torsion scalar
\begin{eqnarray}
&&T= 6H^2 - \dot h_{ij}^2 -4H h_{ij} \dot h_{ij} +a^{-2} \left[ (\partial_i h_{jk})^2 
-(\partial_i h_{jk})(\partial_j h_{ik})\right].
\end{eqnarray}

Then, the second order action for tensor mode is found to be  
\begin{eqnarray}
S_{g,2}^{T}&\eq& \int dt d^3x \ a^3 \left[ -f_0'\dot h_{ij}^2 -4Hf_0' h_{ij} \dot h_{ij} 
+a^{-2}f_0' \left[ (\partial_i h_{jk})^2 -(\partial_i h_{jk})(\partial_j h_{ik}) \right]
-\frac{1}{2}f_0 h_{ij}^2  \right].\nonumber\\
\end{eqnarray}
On the other hand, the second order contribution from the scalar field action is 
\begin{eqnarray}
S_{m,2}^{T}&\eq& \int dt d^3x \ a^3 
\left[-\frac{1}{2}\left(\frac{1}{2}\dot \phi_0^2-V\right)h_{ij}^2\right].
\end{eqnarray}
We can simplify the summation of the above two actions by performing integration by parts, dropping 
the surface terms and using background equation Eq.(\ref{zeta}).

We then have 
\begin{eqnarray}
S_2^T=\int dt d^3x \ a^3 (-f_0') \left[ \dot h_{ij}^2  
-a^{-2} (\partial_i h_{jk})^2 \right].
\end{eqnarray}
As with the scalar perturbation, the only difference of the tensor perturbation from its counterpart in GR is the overall factor.
While in GR the overall factor is 
$1/(2 \kappa)$, 
in $f(T)$ gravity we have the time-dependent coefficient $-f_0'$.

\subsection{Vector and Pseudovector Perturbations}
\label{vector}

Recall that the pseudovector degree of freedom is coupled to the vector degrees of freedom.
Nevertheless by gauge freedom, we can fix the gauge as $\tilde V_i =0 $, so that we need not worry about the pseudovector degrees of freedom. 

The cotetrad and the metric are given by 

\begin{eqnarray}
&&e^0_{\ \mu}dx^\mu= dt + a u_i dx^i,\\ 
&&e^a_{\ \mu} dx^\mu =\delta_{ai}\left[ v_i dt +
a \left(\delta_{ij}+ \partial_i w_j +\partial_j w_i \right) dx^j\right] ,\\
&&e_0^{\ \mu} \frac{\partial}{\partial x^\mu}= (1+u_i v_i) \frac{\partial}{\partial t}
+a^{-1} [ - v_i + v_j (\partial_i w_j +\partial_j w_i )]\frac{\partial}{\partial x^j},\\
&&e_a^{\ \mu}\frac{\partial}{\partial x^\mu}= 
\delta _{ai} \biggl[ \{-u_i + u_j(\partial_i w_j +\partial_j w_i )\} \frac{\partial}{\partial t}
\nonumber\\
&&\qquad
+  a^{-1} \left\{ \delta_{ij} - (\partial_i w_j +\partial_j w_i )+
 (\partial_i w_k +\partial_k w_i )(\partial_j w_k +\partial_k w_j ) +u_i v_j \right\} 
 \frac{\partial}{\partial x^j}\biggr],\\
&&g_{\mu\nu}dx^\mu dx^\nu= -(1-v_i^2) dt^2 
+ 2a[-u_i +v_i + v_j(\partial_i w_j +\partial_j w_i )] dt dx^i  \nonumber\\
&&\qquad
+a^2 \left[\delta_{ij} +2(\partial_i w_j +\partial_j w_i ) -u_i u_j 
+(\partial_i w_k +\partial_k w_i )(\partial_j w_k +\partial_k w_j )\right]dx^i dx^j,\\
&&g^{\mu\nu}\frac{\partial}{\partial x^\mu}\frac{\partial}{\partial x^\nu}=
-(1+2 u_i v_i -u_i^2)  \left(\frac{\partial}{\partial t}\right)^2\nonumber\\
&&\qquad
+2a^{-1} [-u_i+v_i+2u_j(\partial_i w_j +\partial_j w_i )-v_j(\partial_i w_j +\partial_j w_i ) ]
\frac{\partial}{\partial t}\frac{\partial}{\partial x^i}\nonumber\\
&&\qquad
+a^{-2}[\delta_{ij} -2(\partial_i w_j +\partial_j w_i ) +2u_i v_j -v_iv_j \nonumber\\
&&\qquad\qquad\qquad
+3(\partial_i w_k +\partial_k w_i )(\partial_j w_k +\partial_k w_j ) ]
\frac{\partial}{\partial x^i}\frac{\partial}{\partial x^j}.
\end{eqnarray}

We therefore have
\begin{eqnarray}
&&e = a^3 \left[1 -u_i v_i  -\frac{1}{2}(\partial_i w_j +\partial_j w_i )^2 \right].
\end{eqnarray}

The components of the torsion tensor are 
\begin{eqnarray}
&&T^t_{\ ti}= a[\dot u_i -u_j (\partial_i \dot w_j +\partial_j\dot w_i)]  + u_j \partial_i v_j, \\
&&T^t_{\ ij}= a[(\partial_i u_j-\partial_j u_i)-u_k \partial_k (\partial_i w_j -\partial_j w_i)], \\
&&T^i_{\ tj} = H \delta_{ij}+(\partial_i \dot w_j +\partial_j\dot w_i)
-a^{-1}\partial_j v_i -v_i \dot u_j \nonumber\\
&&\qquad\qquad
-(\partial_i  w_k +\partial_k w_i)(\partial_j \dot w_k +\partial_k\dot w_j)  
+ a^{-1}(\partial_i  w_k +\partial_k w_i) \partial_j v_k , \\
&&T^i_{\ jk} =\partial_i(\partial_j w_k - \partial_k w_j)-
v_i (\partial_j u_k - \partial_k u_j)  
- (\partial_i  w_l +\partial_l w_i)\partial_l (\partial_j  w_k -\partial_k w_j),\\
&&T^\mu_{\ t \mu} = 3H -v_i \dot u_i 
-(\partial_i  w_j +\partial_j w_i)(\partial_i \dot w_j +\partial_j\dot w_i)  
+ a^{-1}(\partial_i  w_j +\partial_j w_i) \partial_i v_j ,\\
&&T^\mu_{\ i\mu}= -a \dot u_i -\Delta w_i + a u_j (\partial _i \dot w_j +\partial _j \dot w_i) 
-u_j \partial_i v_j - v_j(\partial_i u_j - \partial_j u_i) \nonumber\\
&&\qquad\qquad
- (\partial_k  w_j +\partial_j w_k)\partial_k (\partial_i  w_j -\partial_j w_i).
\end{eqnarray}

We can compute the various contractions:
\begin{eqnarray} 
&&T_\mu T_\nu g^{\mu\nu}= -9H^2 +\dot u_i^2 + 6H u_i \dot u_i+9H^2 u_i^2 -18H^2 u_i v_i 
+2a^{-1} \dot u_i \Delta w_i \nonumber\\
&&\qquad\qquad\qquad
+6H a^{-1} u_i \Delta w_i 
-6Ha^{-1}\partial_i v_j (\partial_i w_j + \partial_j w_i) -6H a^{-1} v_i \Delta w_i \nonumber\\
&&\qquad\qquad\qquad
+6H(\partial_i w_j +\partial_jw_i)(\partial_i \dot w_j +\partial_j \dot w_i) 
+a^{-2} (\Delta w_i)^2, \\
&&\frac{1}{4}T^\rho_{\ \mu\nu}T^\alpha_{\ \beta\gamma}g_{\rho\alpha}g^{\mu\beta}g^{\nu\gamma}
\nonumber\\
&&\qquad
=-\frac{3}{2}H^2 + \frac{1}{2} \dot u_i^2 + H u_i \dot u_i +\frac{3}{2} H^2 u_i^2 
+\frac{1}{4}a^{-2} (\partial_i u_j - \partial_j u_i)^2 -3 H^2 u_i v_i \nonumber\\
&&\qquad
-\frac{1}{2} a^{-2} (\partial_i v_j)^2 +H a^{-1} u_i \Delta w_i 
+ a^{-1} (\partial_i v_j)( \partial_i \dot w_j +\partial_j \dot w_i)\nonumber\\
&&\qquad
- H a^{-1}(\partial_i v_j)( \partial_i w_j +\partial_j w_i)
-H a^{-1} v_i \Delta w_i 
-\frac{1}{2}( \partial_i \dot w_j +\partial_j \dot w_i)^2\nonumber\\
&&\qquad
+H( \partial_i w_j +\partial_j w_i)( \partial_i \dot w_j +\partial_j \dot w_i)
+\frac{1}{4}a^{-2} [\partial_i( \partial_k w_j +\partial_j w_k)]^2 ,\\
&&\frac{1}{2}T^\rho_{\ \mu\nu}T^\nu_{\ \alpha\rho}g^{\mu\alpha} 
\nonumber\\
&&\qquad
= -\frac{3}{2}H^2 + \frac{1}{2} \dot u_i^2 + H u_i \dot u_i +\frac{3}{2} H^2 u_i^2 
+a^{-2}\partial_i v_j (\partial_i u_j - \partial_j u_i) -3 H^2 u_i v_i \nonumber\\
&&\qquad
-\frac{1}{2} a^{-2} (\partial_i v_j)(\partial_j v_i) +H a^{-1} u_i \Delta w_i 
+ a^{-1} (\partial_i v_j)( \partial_i \dot w_j +\partial_j \dot w_i)
\nonumber\\
&&\qquad
- H a^{-1}(\partial_i v_j)( \partial_i w_j +\partial_j w_i)-H a^{-1} v_i \Delta w_i 
-\frac{1}{2}( \partial_i \dot w_j +\partial_j \dot w_i)^2\nonumber\\
&&\qquad
+H( \partial_i w_j +\partial_j w_i)( \partial_i \dot w_j +\partial_j \dot w_i)\nonumber\\
&&\qquad
+\frac{1}{2}a^{-2} [\partial_i( \partial_k w_j +\partial_j w_k)][\partial_j( \partial_k w_i +\partial_i w_k)].
\end{eqnarray}

Consequently, the torsion scalar is given by 
\begin{eqnarray}
&&T= 6H^2  -4 H u_i \dot u_i -6 H^2 u_i^2 -\frac{1}{4}a^{-2}(\partial_i u_j - \partial_j u_i)^2
+a^{-2}\partial_i v_j (\partial_i u_j - \partial_j u_i) \nonumber\\
&&\qquad
+12 H^2 u_i v_i 
-\frac{1}{2} a^{-2} (\partial_i v_j)^2-\frac{1}{2} a^{-2} (\partial_i v_j)(\partial_j v_i) 
-4H a^{-1} u_i \Delta w_i-2a^{-1} \dot u_i \Delta w_i \nonumber\\
&&\qquad
+2 a^{-1} (\partial_i v_j)( \partial_i \dot w_j +\partial_j \dot w_i)
+4 H a^{-1}(\partial_i v_j)( \partial_i w_j +\partial_j w_i)+4H a^{-1} v_i \Delta w_i \nonumber\\
&&\qquad
-( \partial_i \dot w_j +\partial_j \dot w_i)^2
-4H( \partial_i w_j +\partial_j w_i)( \partial_i \dot w_j +\partial_j \dot w_i)
-a^{-2}(\Delta w_i)^2\nonumber\\
&&\qquad
+\frac{1}{4}a^{-2} [\partial_i( \partial_k w_j +\partial_j w_k)]^2+
\frac{1}{2}a^{-2} [\partial_i( \partial_k w_j +\partial_j w_k)][\partial_j( \partial_k w_i +\partial_i w_k)] .
\end{eqnarray}

Therefore, the second order action for vector mode is given by 
\begin{eqnarray}
S_{g,2}^{V}= \int dt d^3x \ a^3 \left[ f_0 (-u_iv_i +w_i \Delta w_i) +f_0' \delta_2 T \right],
\end{eqnarray}
where $\delta_2 T$ is the second order part of $T$.
The second order part of the scalar field action is
\begin{eqnarray}
S_{m,2}^V= \int dt d^3x \ a^3 \left[
\frac{1}{2}\dot \phi_0^2 (u_i v_i-u_i^2+w_i \Delta w_i) + V(u_i v_i -w_i \Delta w_i)\right].
\end{eqnarray}
We can again simplify the above actions via integrating by parts, dropping 
the surface terms and using background equations Eqs.(\ref{Phi})-(\ref{dotphi}).
The second order total action is finally found to be
\begin{eqnarray}
S_2^V=\int dt d^3x \ \frac{1}{2} a f_0' (v_i -u_i-2a \dot w_i) \Delta (v_i-u_i-2a \dot w_i ).
\end{eqnarray}
From this we can see that up to second order in perturbation, there is no propagating degree of freedom corresponding to vector mode.

\section{Discussion}
\label{Sum}

In conclusion, we examined perturbations of $f(T)$ gravity up to second order and found that there are no propagating scalar, 
pseudoscalar, pseudovector or vector gravitational modes. 
We also found that the scalar and tensor perturbations have exactly the same form as their counterparts in usual general relativity 
with scalar field,
except that the factor $M_{\text{pl}}^2$ that occurs in GR has now been replaced by an effective time-dependent gravitational coupling $-2 (df/dT)|_{T=T_0}$, where $T_0$ is the background torsion scalar.
Our analysis can be applied to TEGR by simply taking $f(T)= -(2\kappa)^{-1}T$ in the analysis above and the result is consistent with known results from TEGR, and hence general relativity. 
Our result is also consistent with the comment in \cite{Li} that the extra degrees of freedom do not appear in Minkowski spacetime. 
This can be seen by taking the limit $H \to 0$\footnote{ 
In the case of scalar modes, the second order action (\ref{scalarsecond}) seems to diverge in this limit due to the factor $(\dot H + 3H^2)^{-1}$, but after expanding the brackets we can easily check that it is finite and that there are no extra degrees of freedom.
}.

We nevertheless emphasize that our current calculation only reveals dynamics of the propagating modes in $f(T)$ gravity in a \emph{highly symmetric} background of FLRW geometry. Although this work is useful for the purpose of studying some aspects of cosmological properties, it does not reveal the full features of what $f(T)$ theory has to offer. In general $f(T)$ gravity has 
three extra degrees of freedom 
in nonlinear analysis
\cite{Li}, and the extra degrees of freedom might be excited in a spacetime with \emph{less} symmetry. 

The situation is very similar to that in nonlinear massive gravity theory \cite{deRham:2010kj}
in which 
nonlinear analysis shows that there are generically five gravitational degrees of freedom \cite{Hassan:2011hr}.
There are in fact \emph{three} branches of solutions describing open FLRW universe \cite{Gumrukcuoglu:2011ew}. 
On one of them all five degrees of freedom appear in linear analysis. 
However on the other two branches, the second order action 
only displays two tensor degrees of freedom \cite{Gumrukcuoglu:2011zh}, 
which is a situation similar to our analysis.
Since the hidden degrees of freedom cause nonlinear instability in the case of nonlinear massive gravity \cite{Shinji,DeFelice:2012mx},
we cannot conclusively decide on the issue of stability in $f(T)$ theory until we know the behavior of all the degrees of freedom that might manifest under more generic conditions. This is an important issue in cosmology since the real universe is certainly \emph{not} completely homogeneous and isotropic.
 
A fully nonlinear analysis for $f(T)$ gravity is expected to be difficult to perform. It thus remains an interesting problem to seek the effects of these degrees of freedom at \emph{linear} perturbation level in an anisotropic model of cosmology \cite{RHSR}, e.g. Bianchi Type-I geometry \cite{Sharif}, since such perturbation can be viewed as \emph{effectively} \emph{nonlinear} perturbation on FLRW cosmology. For such analysis in the context of nonlinear massive gravity, see e.g. \cite{DeFelice:2012mx}. We will address these interesting issues elsewhere.

\acknowledgments
The authors would like to thank James Nester for valuable discussions about teleparallel theories, Je-An Gu for useful discussions about $f(T)$ gravity, and Lance Labun for proof-reading the manuscript. The authors are also grateful to 
Pisin Chen for much appreciated help and various supports.
Keisuke Izumi is also grateful to Shinji Mukohyama for useful discussions and comments.  
In addition, 
Yen Chin Ong would like to thank Brett McInnes
for useful conversation about torsion, as well as Wu-Hsing Huang for fruitful discussions about various aspects of differential geometry. 
Keisuke Izumi is supported by Taiwan National Science Council under Project No. NSC101-2811-M-002-103. Yen Chin Ong is supported by the Taiwan Scholarship from Taiwan's Ministry of Education.

\end{document}